\documentclass[a4paper,fleqn,usenatbib]{mnras}

\usepackage[T1]{fontenc}
\usepackage{ae,aecompl}

\usepackage{graphicx}	
\usepackage{amsmath}	
\usepackage{mathptmx}
\usepackage{mathptmx}
\usepackage{txfonts}
\usepackage{amssymb}	

\title[Chemical evolution of M\,101]
{Breaking the degeneracy between gas inflow and outflows with stellar metallicity:
Insights on M\,101}

\author[X. Y. Kang et al. ]
{Xiaoyu~Kang$^{1,2,3}$\thanks{E-mail: kxyysl@ynao.ac.cn}, Ruixiang~Chang$^{4}$, Rolf-Peter~Kudritzki$^{5,6}$,
Xiaobo~Gong$^{1,2,3,7}$ \and and Fenghui~Zhang$^{1,2,3}$ \\
$^1$Yunnan Observatories, Chinese Academy of Sciences, 396 Yangfangwang, Guandu District, Kunming, 650216, P.R. China\\
$^2$Key Laboratory for the Structure and Evolution of Celestial Objects, Chinese Academy of Sciences, 396 Yangfangwang,\\
Guandu District, Kunming, 650216, P. R. China\\
$^3$Center for Astronomical Mega-Science, Chinese Academy of Sciences, 20A Datun Road, Chaoyang District, Beijing,\\ 100012, P.\,R.\,China\\
$^4$Key Laboratory for Research in Galaxies and Cosmology, Shanghai Astronomical Observatory, Chinese Academy of \\
Sciences, 80 Nandan Road, Shanghai, 200030, China\\
$^5$LMU M$\rm \ddot{u}$nchen, Universit$\rm \ddot{a}$tssternwarte, Scheinerstr. 1, 81679 M$\rm \ddot{u}$nchen, Germany\\
$^6$Institute for Astronomy, University of Hawaii, 2680 Woodlawn Drive, Honolulu, HI96822, USA\\
$^7$University of Chinese Academy of Sciences, Beijing, 100049, P.\,R.\,China
}

\date{Accepted XXX. Received YYY; in original form ZZZ}

\pubyear{2020}

\begin{document}
\label{firstpage}
\pagerange{\pageref{firstpage}--\pageref{lastpage}}
\maketitle

\begin{abstract}

An analytical chemical evolution model is constructed to investigate the radial
distribution of gas-phase and stellar metallicity for star-forming galaxies.
By means of the model, the gas-phase and stellar metallicity can be obtained from the
stellar-to-gas mass ratio. Both the gas inflow and outflow processes play an important
role in building the final gas-phase metallicity, and there exists degeneracy effect between
the gas inflow and outflow rates for star-forming galaxies. On the other hand, stellar
metallicity is more sensitive to the gas outflow rate than to the gas inflow rate,
and this helps to break the parameter degeneracy for star-forming galaxies.
We apply this analysis method to the nearby
disc galaxy M\,101 and adopting
the classical $\chi^{2}$ methodology to explore the influence of
model parameters on the resulted metallicity. It can be found
that the combination of gas-phase and stellar metallicity is indeed more
effective for constraining the gas inflow and outflow rates.
Our results also show that the model with relatively strong gas outflows
but weak gas inflow describes the evolution of M\,101 reasonably well.
\end{abstract}

\begin{keywords}
galaxies: evolution -- galaxies: abundances -- galaxies: stellar content --
galaxies: individual (M\,101) -- galaxies: spiral
\end{keywords}



\section{Introduction}
\label{sec:intro}

Chemical evolution modelling is a powerful tool to explore
the galactic formation and evolution. Analytical chemical
evolution models, based on simple parametrisation of key
physical processes, such as gas accretion, star formation, nucleosynthesis,
stellar mass return, and gas outflows, have been successfully employed to predict
the enrichment of the interstellar medium (ISM) and achieved
a series of interesting results. In the Milky Way, the closed-boxed
chemical evolution model predicts a higher fraction of metal-poor G-dwarf
stars in the solar neighborhood than observed (the classical "G-dwarf problem"),
suggesting that the Milky Way disc is not a closed-boxed system and gas inflow
is important for galactic evolution \citep[][and references therein]{Chang1999}.

The average chemical composition of the stars and the ISM can
both provide constraints on the chemical enrichment history of galaxies.
The gas-phase metallicity provides a snapshot of the metal content
at a given time, while the mean stellar metallicity reflects the time-averaged
value of the ISM metal content over the star formation history (SFH) of the galaxy.
Most analytical chemical evolution studies in the literature focused
on modelling gas-phase metallicity, since it is easier to measure than stellar
metallicity in systems where individual stars cannot be resolved.
Early analytical models derived a relation between gas-phase
metallicity and the ratio of stellar to gas mass \citep{Zahid2014,Yabe2015,Kudritzki2015}.
However, degeneracy among the model parameters (star formation
efficiency (SFE, with the definition of the proportion of gas turns into
stellar mass in unit time), the gas inflow and outflow rates
(their definitions are in Section\,\ref{sec:model})), leads to severe
limitations when using the gas-phase metallicity alone to constrain the model
\citep{Kudritzki2015, Belfiore2016}.
Better constraints can be obtained if both the stellar
and the gas-phase metallicity are used at the same time.
Fortunately, as demonstrated, for instance, by \citet{Zahid2017},
the stellar metallicity of star-forming galaxies can be obtained from a
population synthesis analysis of spectra of integrated stellar populations.
Large samples of galaxies observed with spatially resolved spectroscopy are
becoming available through recent integral field unit (IFU) surveys like CALIFA
\citep{Sanchez2012} and MaNGA \citep{Bundy2015}.
Exploiting early data from such integral field spectroscopy surveys,
\citet{Lian2018a, Lian2018b} simultaneously explored the gas-phase and
stellar metallicity, and pointed out that the stellar metallicity may
contribute to breaking the degeneracy between parameters in analytical
chemical evolution models.

To further investigate the importance of stellar metallicity in chemical evolution
studies we take NGC\,5457 (M\,101) as an example and focus on its radial
metallicity gradient. M\,101 is a nearby face-on Scd galaxy \citep{Freedman2001},
which is known to be currently experiencing an inflow of high-velocity gas
\citep{Sancisi2008}, and has likely been recently subjected to interaction events
\citep{Waller1997, Mihos2012}. Its basic observational properties are summarized in
Table\,\ref{Tab:obs1},
and their corresponding values are taken from \citet{Walter2008}.
Observations for gas-phase metallicity of H{\sc ii}
regions along the disc of M\,101 have been carried out since the 1970s
\citep{Searle1971, Smith1975, McCall1985, Kennicutt1996, Kennicutt2003,
Bresolin2007, LiYX2013, Croxall2016, Huning2018, Esteban2020}.
\citet{Croxall2016} carried out the most extensive study to date of
oxygen abundance in M101, deriving a metallicity gradient with slope
$-0.027\,\pm\,0.001\,\rm dex\,kpc^{-1}$.
\citet{Linlin2013} derived the stellar metallicity gradient of M101,
using spectral energy distribution (SED) fitting of ultraviolet, optical
and infrared photometry. They found the stellar metallicity gradient to
be flatter than that of H{\sc ii} regions. However, model explanation of
these observed properties is still lacking.

The aim of this work is to investigate whether the analytical model
introduced in Section\,\ref{sec:model} can explain the radial
distributions of both gas-phase and stellar metallicity,
and whether the stellar metallicity can help to relieve the
degeneracy between gas inflow and outflows.
The structure of this paper is as follows. The main ingredients of
the model are described in Section\,\ref{sec:model}. The observations are presented in
Section\,\ref{sec:observe}. Our main results and discussion are shown in
Section\,\ref{sec:result}.
The last section summarizes our main conclusions.

\begin{table}
\caption{Basic properties of M\,101.}
\label{Tab:obs1}
\begin{center}
\begin{tabular}{lc}
\hline
Property            &    Value                             \\
\hline
name                &    M\,101, NGC\,5457                      \\
RA                  &    $14^{\rm h}03^{\rm m}12^{\rm s}.6$     \\
Dec                 &    $+54^{\rm \circ}20^{\rm '}57^{\rm ''}$ \\
Morphology          &    Scd                       \\
Distance(adopted)   &    $7.4\,\rm Mpc$               \\
Inclination         &    $18^{\circ}$                                  \\
$R_{25}$            &    $25.81\,\rm kpc$                       \\
Scale               &    $36\rm pc\,arcsec^{-1} $               \\
\hline
\end{tabular}\\
\end{center}
\end{table}

\section{The model}
\label{sec:model}
\begin{figure*}
  \centering
  \includegraphics[angle=0,scale=0.7]{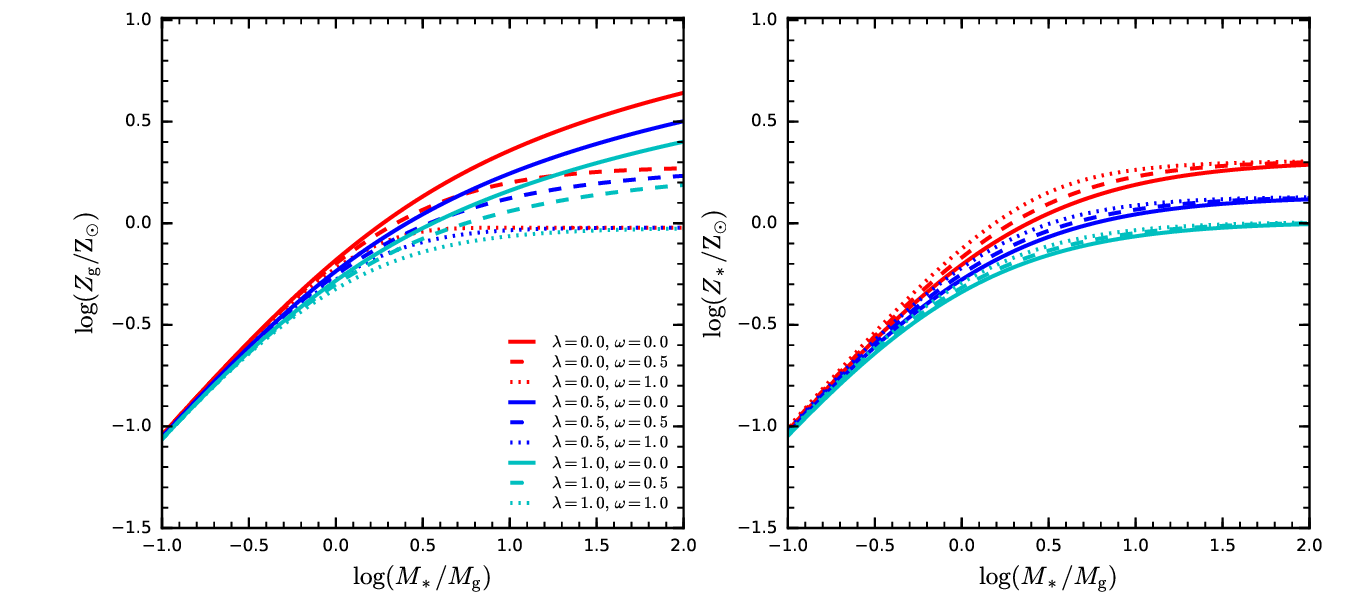}
    \caption{Metallicity as a function of the stellar-to-gas mass
      ratio with different combinations of
gas outflow loading factor ($\lambda$) and gas mass accretion factor ($\omega$).
The lines with different color represent different values of $\lambda$, while
the different line-types are corresponding to different values of $\omega$.
Gas-phase metallicity and stellar metallicity are displayed in the left panel and in
the right panel, respectively.
}
\label{fig:f0}
\end{figure*}
Similar to our previous work \citep{Chang1999, Kang2016, Kang2017}, we assume that
a star-forming galaxy is gradually built up due to
continuous gas inflow. At the same time, outflows of metal
enriched gas are also taken into account. We adopt the instantaneous recycling
approximation (IRA) assuming that the gas return from stars to the ISM happens
on a short timescale compared with galactic evolution, and we assume
that the gas is well mixed with stellar ejecta.
IRA represents a good approximation for oxygen produced by massive stars
with short lifetimes.
The chemical evolution of a galaxy is expressed
by the classical set of integro-differential equations from \citet{1980FCPh....5..287T}:
\begin{equation}
\frac{{\rm d}M_{\rm g}}{{\rm d}t}\,=\,\Phi-\Psi-(1-R)\rm SFR,
\label{eq:gas}
\end{equation}
\begin{equation}
\frac{{\rm d}M_{\ast}}{{\rm d}t}\,=\,(1-R)\rm SFR,
\label{eq:star}
\end{equation}
\begin{equation}
\frac{{\rm d}[Z_{\rm g}\cdot M_{\rm g}]}{{\rm d}t}\,=\,y(1-R){\rm SFR}-Z_{\rm g}(1-R){\rm SFR}
+Z_{\rm{i}}\Phi-Z_{\rm{o}}\Psi,
\label{eq:metallicity}
\end{equation}
where $M_{\rm g}$ and  $M_{*}$ are the gas mass and the stellar mass of the galaxy at
evolution time $t$, respectively; $Z_{\rm g}$ is the gas-phase metallicity of the system.
$\Phi$ and $\Psi$ are the gas inflow rate and the gas outflow rate, respectively.
SFR is the star formation rate (SFR).
$R$ is the return mass fraction and $y$ is the nucleosynthesis yield.
Both $R$ and $y$ depend on the adopted stellar
initial mass function (IMF) but are only weakly dependent on metallicity and time
\citep{Vincenzo2016}. The IMF of \citet{Kroupa1993} is adopted in this work, since this IMF is favored
in describing the chemical evolution of the disc of spirals similar to the Milky Way
\citep{Vincenzo2016}. Neither metallicity nor time dependence of $R$ and $y$
will be further taken into account in our model.
The values of $R$ and $y$
are taken from Table\,2 of \citet{Vincenzo2016} corresponding to the stellar yields
of \citet{Romano2010}. We obtain $R\,=\,0.289$ and $y\,=\,0.019$ from averaging the
values of $R$ and $y_{Z}$ over metallicity, respectively.
$Z_{\rm{i}}$ is the inflowing gas-phase metallicity. We adopted metal-free inflow,
i.e., $Z_{\rm{i}}=0$. $Z_{\rm{o}}$ is the outflowing gas-phase metallicity and
assumed to have the same metallicity as the ISM,
i.e., $Z_{\rm{o}}=Z_{\rm g}$ \citep{Chang2010,Kang2012,Kang2016,Kang2017}.

Our model requires an assumption about matter inflow
and outflows. Many chemical evolution models adopt an inflow rate of gas which
follows an exponential law \citep[][and references therein]{Matteucci1989,
Hou2000, Spitoni2017}. An alternative, which we adopt for our model, is to assume
that the gas inflow rate is proportional to the SFR, since the inflow of gas
provides a continuous reservoir for star formation. Physical arguments in
support of this assumption can be found in \citet{Matteucci1983}, \citet{Recchi2008},
\citet{Bouche2010}, \citet{Lilly2013} and \citet{Yabe2015}.
For the outflow rate we also assume that it is also proportional to the SFR because the larger is the SFR,
the larger is the chance of having a larger-scale outflow
\citep{Silk2003}.
With these assumptions analytical solutions of chemical evolution are
straightforward \citep{Recchi2008, Spitoni2010, Kudritzki2015}. We use
$\Phi\,=\,\omega(1-R)\rm SFR$,
$\Psi\,=\,\lambda(1-R)\rm SFR$, where $\omega$ ($\omega\geq0$) and $\lambda$ ($\lambda\geq0$)
are the gas mass accretion factor and the outflow loading factor, respectively.
$\omega$ and $\lambda$ are two free parameters in our model.
It should be pointed out that, under these assumptions, equation
(\ref{eq:gas}) combined with a linear star formation law from
\citet{Schmidt1959} (i.e., $\rm SFR\,=\,\varepsilon M_{\rm gas}$, $\varepsilon$
is the so-called SFE in units of $\rm Gyr^{-1}$) will lead to
an exponentially declining SFR, that is, the inflow rate obeys the decaying
exponential law, in agreement with the approach in \citet{Matteucci1989}.

With the definition $\alpha=\lambda-\omega$, equation (\ref{eq:gas}) and
(\ref{eq:metallicity}) can be re-written as follows:
\begin{eqnarray}
{\rm d}M_{\rm g}\,=\,-(1+\alpha)\cdot{\rm d}M_{\ast},
\label{eq:gas1}\\
{\rm d}[Z_{\rm g}\cdot M_{\rm g}]\,=\,(y-Z_{\rm g}-pZ_{\rm g})\cdot{\rm d}M_{\ast},
\label{eq:metallicity1}
\end{eqnarray}

In this paper, we define the mean stellar metallicity as the mass-weighted average stellar metallicity,
\begin{equation}
\langle\,Z_{\ast}\rangle\,=\,\frac{(1-R)\int^{t}_{0} Z_{\rm g}(t'){\rm SFR}(t'){\rm d}t'}{(1-R)\int^{t}_{0} {\rm SFR}(t'){\rm d}t'}.
\label{eq:Zstar}
\end{equation}
Integrating equations (\ref{eq:gas1}) and (\ref{eq:metallicity1}), and
combining with equation (\ref{eq:Zstar}), we can obtain the
analytical solutions of the chemical evolution
using appropriate initial conditions.
In other words, both the gas-phase metallicity and the mean stellar metallicity
solutions can be obtained.
There are four different solutions corresponding to four special cases.

The first case includes both gas inflow and outflows, i.e., $\omega\,\neq\,0$ and
$\lambda\,\neq\,0$, and $\alpha\,\neq\,-1$. The initial conditions are
$M_{\ast,0}\,=\,0$ and $M_{\rm g,0}\,=\,M_{\rm g}+(1+\alpha)M_{\ast}$, and the solutions are
\begin{equation}
\left\{\begin{array}{ll}
            Z_{\rm g}\,=\,\frac{y}{\omega}\{1-[1+(1+\alpha)\frac{M_{*}}{M_{\rm g}}]^{-\beta}\},\\
            \langle\,Z_{\rm\ast}\rangle\,=\,-\frac{y}{\omega(1+\alpha)}\{[1-\frac{1+\alpha}{1+\lambda}[1+(1+\alpha)\frac{M_{*}}{M_{\rm g}}]^{-\beta}]\frac{M_{\rm g}}{M_{\ast}}\\
            -\frac{\omega}{1+\lambda}[\frac{M_{\rm g}}{M_{\ast}}+(1+\alpha)]\}.
           \end{array}
       \right.
\label{eq:metallicityboth}
\end{equation}
where $\beta=\frac{\omega}{1+\alpha}$. We should emphasize that the values of
$\lambda$ and $\omega$ are conditional.
Since the initial gas mass should not be less than zero, i.e., $M_{\rm g,0}\,\geq\,0$, the value of
mass accretion factor is constrained to the range $\omega\,\leq\,\lambda+\frac{1}{1-\mu}$, where $\mu$ is the
gas fraction and defined as $\mu\,=\,\frac{M_{\rm g}}{M_{\rm g}+M_{\ast}}$.

The second case is $\alpha\,=\,-1$. The initial
conditions are $M_{\ast,0}\,=\,0$ and $M_{\rm g,0}\,=\,M_{\rm g}\,=\,\rm const$,
and the corresponding solutions are
\begin{equation}
\left\{ \begin{array}{ll}
           Z_{\rm g}\,=\,\frac{y}{\omega}(1-e^{-\omega\frac{M_{*}}{M_{\rm g}}}),\\
           \langle\,Z_{\rm \ast}\rangle\,=\,\frac{y}{\omega}[1+\frac{M_{\rm g}}{\omega M_{*}}(e^{-\omega\frac{M_{*}}{M_{\rm g}}}-1)].
           \end{array}
       \right.
\label{eq:metallicityboth1}
\end{equation}

The third case is no gas inflow but with gas outflow, i.e.,
$\omega\,=\,0$ and $\lambda\,\neq\,0$. The initial conditions are $M_{\ast,0}\,=\,0$
and $M_{\rm g,0}\,=\,M_{\rm g}+(1+\lambda)M_{\ast}$, and the solutions are
\begin{equation}
\left\{ \begin{array}{ll}
          Z_{\rm g}\,=\,\frac{y}{1+\lambda}{\rm ln}[1+(1+\lambda)\frac{M_{*}}{M_{\rm g}}],\\
          \langle\,Z_{\ast}\rangle\,=\,\frac{y}{(1+\lambda)^{2}}\{(1+\lambda)-\frac{M_{\rm g}}{M_{*}}{\rm ln}[1+(1+\lambda)\frac{M_{*}}{M_{\rm g}}]\}.
           \end{array}
       \right.
\label{eq:metallicityoninfall}
\end{equation}

The last one is the closed-box model with neither gas inflows nor outflows, i.e., $\omega\,=\,0$
and $\lambda\,=\,0$, the initial conditions are $M_{\ast,0}\,=\,0$
and $M_{\rm g,0}\,=\,M_{\rm g}+M_{\ast}$, and the solutions
become
\begin{equation}
\left\{ \begin{array}{ll}
          Z_{\rm g}\,=\,y{\rm ln}[1+\frac{M_{*}}{M_{\rm g}}],\\
          \langle\,Z_{\ast}\rangle\,=\,y[1-\frac{M_{\rm g}}{M_{*}}{\rm ln}(1+\frac{M_{*}}{M_{\rm g}})].
           \end{array}
       \right.
\label{eq:closedbox}
\end{equation}

The above equations show that, given the two free parameters $\lambda$
and $\omega$, the gas-phase and stellar metallicity at time $t$ can be calculated
by the stellar-to-gas mass ratio at that moment. In order to illustrate this
point clearly, Figure\,\ref{fig:f0} displays the gas-phase metallicity (left)
and the stellar metallicity (right) as a function of the stellar-to-gas mass
ratio with different combinations of the parameters $\lambda$
    and $\omega$. The lines with different colors correspond
to different outflow loading factors (i.e., red $\lambda\,=\,0$, blue $\lambda\,=\,0.5$ and cyan
$\lambda\,=\,1.0$), while the different line-types are corresponding to different mass accretion
factors (i.e., solid $\omega\,=\,0$, dashed $\omega\,=\,0.5$ and dotted $\omega\,=\,1.0$).
We should point out that the left panel of Figure\,\ref{fig:f0} is similar to the
Figure\,1 of \citet{Kudritzki2015}. Here we add the panel of stellar metallicity as a function
of the stellar-to-gas mass ratio and simultaneously explore the gas-phase and stellar metallicity.


The left panel of Figure\,\ref{fig:f0} shows that the closed-box model (the red-solid
line) provides the upper limits of the gas-phase metallicity. Gas outflow removes
part of metal content from the system and reduces the final gas-phase metallicity,
while gas inflow slows down the ISM chemical enrichment by adding
pristine gas. Unfortunately, since both the gas inflow and the gas outflows
reduce the gas-phase metallicity, a degeneracy between parameters $\lambda$
and $\omega$ exists in the stellar-to-gas mass range of
${\rm log}(M_{*}/M_{\rm g})\,=\,0$ to ${\rm log}(M_{*}/M_{\rm g})\,=\,1.4$,
which makes it difficult to disentangle the role of outflow
and inflow from the observations of the gas-phase metallicity only.

On the other hand, the stellar component in the right panel
of Figure\,\ref{fig:f0} shows a different behavior for
$-0.2\,\leq\,{\rm log}(M_{*}/M_{\rm g})\,\leq\,1.4$. For a given outflow loading
factor $\lambda$, the mean stellar metallicity increases with the mass accretion factor $\omega$,
because large inflow factor means large fraction of star forms at late time and then
having high metallicity. Based on the gas-to-stellar relation of
\citet[][Eq.(9)]{Peeples2014} from $\sim260$ star-forming galaxies,
the typical range of the stellar-to-gas ratio is about
$-0.07\,\leq\,{\rm log}(M_{*}/M_{\rm g})\,\leq\,1.13$ for star-forming galaxies
with stellar mass $9.0\,\leq\,{\rm log}(M_{*}/M_{\odot})\,\leq\,11.5$.
Therefore, the observed stellar metallicity may help us to
overcome the parameter degeneracy and serve as an important observable to constrain
the SFHs of star-forming galaxies. In the following
Sections, we choose M\,101 as an example to demonstrate the importance of stellar
metallicity in the chemical enrichment studies.

\begin{figure*}
  \centering
  \includegraphics[angle=0,scale=0.7]{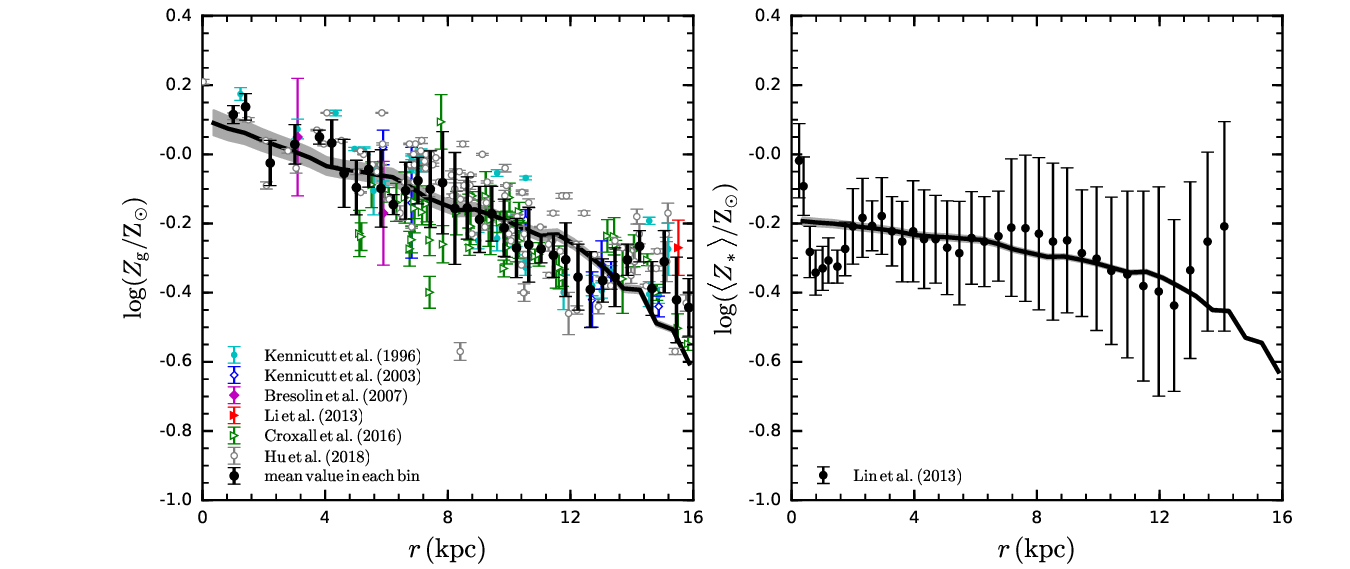}
\caption{Comparison of model predicted metallicity with
observations in M\,101 (left: gas-phase; right: stars).
\emph{Left panel}: different symbols denote the observed gas-phase metallicity
from different authors. The cyan solid and grey open cycles
represent the gas-phase metallicity calculated by using the calibration of
KK04, while the blue open diamonds, magenta solid
diamonds, the red solid triangles and green open triangles are corresponding to
the gas-phase metallicity derived by using the direct \emph{Te} calibrations.
Note that the KK04 calibration data are corrected to remove the discrepancies
between \emph{Te} and KK04 calibrations (see text). The black solid cycles display
the mean values of observed gas-phase metallicity in each bin, and the error
bars are the standard deviation of data in each bin. \emph{Right panel}: the observed
stellar metallicity data are shown as black solid cycles.
The solid lines in both panels plot the best-fitting model predictions
adopting ($\lambda\,=\,1.137, \omega\,=\,0.781$), while the grey shaded area denote
the model predictions adopting ($\lambda\,=\,1.069, \omega\,=\,0.689$) and
($\lambda\,=\,1.207, \omega\,=\,0.869$) with $1\sigma$ confidence level.}
\label{fig:f1}
\end{figure*}

\section{The observations}
\label{sec:observe}

As described in Section\,\ref{sec:model}, with our model, spatially resolved
gas-phase and stellar metallicities can be calculated from the spatially resolved
stellar-to-gas mass ratio. In this Section, we summarize the observed radial
distribution of these properties for M\,101, including the gas-phase metallicity,
the stellar metallicity, the mass surface densities of ISM neutral and molecular
hydrogen and the stellar mass surface density.

\subsection{Stellar and gas mass surface densities}
\label{sec:gasstar}

The stellar mass surface density ($\Sigma_*$) is derived
from infrared (IR) surface
photometry obtained with the 2MASS survey \citep{Jarrett2003}
using the \emph{K}-band at $2.2\,\rm \mu m$. The \emph{K}-band surface brightness
profile from \citet{MM2007} and the fixed \emph{K}-band mass-to-light ratio,
$\Upsilon_{\ast}^{K}\,=\,0.5\,{\rm M}_{\odot}/{\rm L}_{\odot,K}$ \citep[see][]{Leroy2008},
is adopted to calculate $\Sigma_*$.

The neutral hydrogen gas mass surface density ($\Sigma_{\rm HI}$) of M\,101 is obtained from
Very Large Array (VLA) maps of the 21-cm hydrogen line as part of The H{\sc i} Nearby
Galaxy Survey \citep[THINGS;][]{Walter2008}. The molecular hydrogen gas mass surface
density ($\Sigma_{\rm H_{2}}$) is derived by CO ($J=2-1$) maps carried out by using the
IRAM $\rm 30\,m$ as part of the HERA CO-Line Extragalactic Survey \citep[HERACLES;][]{Leroy2009}.
A factor of 1.36 has been included to account for the contribution of helium and heavier
elements, and the reader is referred to \citet{Schruba2011} for more details about
the conversion of emission (21-cm and CO) line into $\Sigma_{\rm HI}$ and $\Sigma_{\rm H_{2}}$,
respectively. The total gas mass surface density ($\Sigma_{\rm gas}$) is
defined as $\Sigma_{\rm gas}\,=\,\Sigma_{\rm HI}\,+\,\Sigma_{\rm H_{2}}$.

\subsection{Metallicity gradients}
\label{sec:metal}

Since oxygen is the element most commonly measured and taken as a
tracer for the total metal content, and because it is an element for which
the IRA approximation is appropriate, we will use the oxygen abundance to
represent the metallicity of M\,101 and adopt the solar value as
$\rm 12+log(O/H)_{\odot}\,=\,8.69$ \citep{Asplund2009} throughout this work.

The radial distribution of gas-phase metallicity of H{\sc ii} regions in M\,101
have been obtained in several works, notably \citet{Kennicutt1996}, \citet{Kennicutt2003},
\citet{Bresolin2007}, \citet{LiYX2013} and \citet{Croxall2016} and \citet{Huning2018}.
The gas-phase metallicity from \citet{Kennicutt1996} and \citet{Huning2018} are
calculated by using the theoretical calibration published
by \citet[][hereafter KK04]{KK04}, while those from \citet{Kennicutt2003},
\citet{Bresolin2007}, \citet{LiYX2013} and \citet{Croxall2016} are derived by
using the direct \emph{T}e methods.
Since the absolute gas-phase metallicity depends on
the calibrations used, it is crucial to use the
same metallicity calibration when using the observations to constrain the model.
\emph{T}e method is considered as the most reliable approach to determine the
gas-phase matallicity \citep{Izotov2006}, and the gas-phase
metallicity obtained from the direct \emph{T}e method is systemically
$\sim0.4\,\rm dex$ lower than that from the KK04 calibration
\citep{Huning2018}. Consequently, we use the gas-phase metallicity
data obtained from direct \emph{T}e calibration
to constrain the model. We also add the data based on the KK04 but
subtract 0.4 dex to account for the systematic effect of this calibration.

The radial distribution of stellar metallicity for M\,101 is derived
by \citet{Linlin2013}, who fitted evolutionary population synthesis model \citep{BC2003}
to a set of multi-band photometry images from ultraviolet, optical and
infrared together with the 15 intermediate-band images observed in the
Beijing-Arizona-Taiwan-Connecticut (BATC) filter system.
The BATC photometric system covers the wavelength range of $3300-10000\,{\AA}$,
and the bandwidths of the intermediate-band filters are about $200-300\,{\AA}$.
The reader is referred to Table\,1 in \citet{Linlin2013} for more details about
the effective wavelengths of the filters and some statistics of the stacked images.
It should be pointed out that \citet{Linlin2013} used a fixed metallicity
when performing SED fitting for each pixel, that is, simple stellar populations of different ages
have the same metallicity. Thus, the mean stellar metallicity they obtained
inclines to a luminosity-weighted metallicity. The luminosity-weighted metallicity
cannot be corrected to the mass-weighted metallicity by using the mass-to-light ratio
in a specific band. Moreover, since the simple model we are
using does not contain specific information of SFH,
we are not able to convert a mass-weighted metallicity
predicted by our model into a luminosity-weighted metallicity.
We have to tolerate that this may add some uncertainties to our results.
Fortunately, Figure\,11 in \citet{Linlin2013} shows that the disc of M\,101
is dominated by intermediate-age stellar populations ($\sim6\,\rm Gyr$)
and the age gradient is quite flat, which indicates the mass-to-light ratio
may not vary significantly in the disc of M\,101.
It should also be noted that the stellar metallicity data in \citet{Linlin2013}
are obtained from multi-band photometry. Such a determination is generally
very uncertain, and this is a challenge for star-forming galaxies. To use
spectroscopic stellar metallicity to constrain the model would be
much better, unfortunately, no such data are available for M\,101.

The radial distribution of metallicity depends on the distance to M\,101,
and we note that the distance used in \citet{Schruba2011} is the
same as the distances used by
\citet{Linlin2013}, \citet{Croxall2016} and \citet{Huning2018}, but different from
those used in \citet{Kennicutt1996}, \citet{Kennicutt2003}, \citet{Bresolin2007} and
\citet{LiYX2013}. Consequently, we have scaled all the metallicity distributions of M\,101 to the
distance used by \citet{Schruba2011},
and the value of the distance to M\,101 is taken from \citet{Karachentsev2004}.
The left panel of Figure\,\ref{fig:f1} plots the observed radial distribution of
gas-phase metallicity of H{\sc ii} regions in M\,101 from different authors as
different symbols, and the observed radial distribution of stellar metallicity data
from \citet{Linlin2013} is displayed in the right panel of Figure\,\ref{fig:f1}.
It should be pointed out that, since the gas mass
surface density taken from \citet{Schruba2011} goes only out to $r\sim16\,\rm kpc$,
we only adopt the observed
gas-phase metallicity within the disc range $r\,\leq\,16\,\rm kpc$.

The solid lines of Figure\,\ref{fig:f1} are the predictions of the
best-fitting model, and the grey shaded regions are the model
predictions that enclose $1\sigma$ confidence level,
which will be described in detail in the following Section.

\begin{figure*}
  \includegraphics[angle=0,scale=0.7]{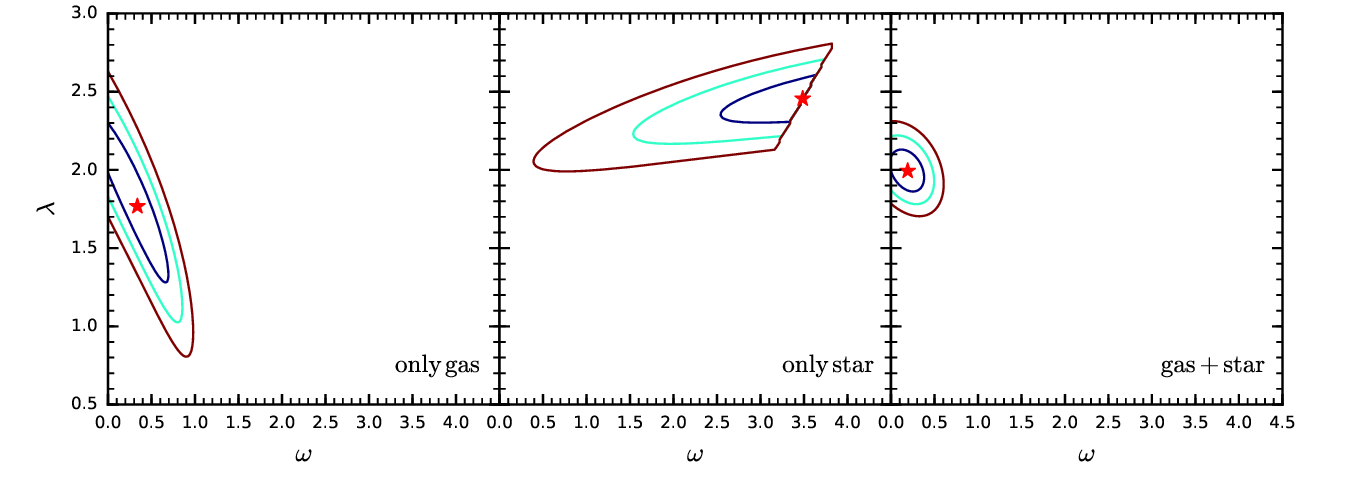}
\caption{$\chi^{2}$ contour maps for the outflow loading factor
$\lambda$ and the mass accretion $\omega$ determinations.
Only gas-phase metallicity (left), only stellar metallicity (middle)
and both gas-phase and stellar metallicity (right) are used to
constrain the model.
The red filled asterisk in each panel denotes the minimum value of
$\chi^{2}$. The $\Delta\chi^{2}$ values adopted for these plots
are 2.3 (blue), 6.17 (cyan) and 11.8 (brown), which are corresponding to
$68.3\%$, $95.4\%$ and $99.73\%$ confidential levels, respectively.
}
\label{fig:f2}
\end{figure*}

\section{results and discussion}
\label{sec:result}


As has been described in Sections \ref{sec:model} and \ref{sec:observe},
with the observed radial distributions of the stellar mass, gas mass and metallicity
in the disc of M\,101 together with the metallicity yield $y$, we can determine
the outflow loading factor $\lambda$ and the mass accretion factor $\omega$
of the analytical chemical evolution model presented in Section \ref{sec:model}.
For a given galactic-centric radius $r$, we use the observed stellar-to-gas ratio
at the present-day to calculate gas-phase and stellar metallicity based on the model
solutions in Section\,\ref{sec:model}.
In other words, for each parameter combination of $\lambda$ and $\omega$, we generate
a model library of radial metallicity profile
for both gas-phase and stellar components. We use the classical $\chi^{2}$ methodology to compare
the model predictions with the corresponding observed
data. Refering to the definition by \citet[][Eq.(15.1.5)]{Press1992}, we adopt
$\chi^{2}\,=\,\sum\limits_{i=1}^{N}\frac{(C_{{\rm model},i}-C_{{\rm obs},i})^{2}}{\sigma_{i}^{2}}$,
where $\sigma_{i}$ is the observed error and $N$ is the
number of observed data. The model that minimizes this reduced $\chi^{2}$ is considered as
the best-fitting model.

In order to ensure gas-phase and stellar metallicity have nearly the
same weight, we divide the radial observed gas-phase metallicity data introduced above along
the disc ($r\,\leq\,16\,\rm kpc$) of M\,101 into 36 bins, which is the same number as that of the
stellar component. Then, we calculate the mean values of gas-phase metallicity in each bin and
show them as big black solid cycles in the left panel of Figure\,\ref{fig:f1}, where
the error bars represent the standard deviation of data in each bin.
Since the error of some gas-phase metallicity data are
not known, following the approach of \citet[][Eq.(15.1.6)]{Press1992},
we assume that all the observed metallicity data have the same standard
deviation, $\sigma_{i}\,=\,\sigma$, and that the model does fit well, that is,
we adopt the reduced $\chi^{2}$, $\chi^{2}_{\nu}\,=\,1$, to fit the data.
According to $\chi^{2}_{\nu}\,=\,\frac{\chi^{2}}{N-M}$, where $N-M$ is the
number of degrees of freedom for fitting $N$ data points with $M$ parameters,
we can get $\chi^{2}\,=\,N-M$.

We separately calculate the values of $\chi^{2}$ for gas-phase metallicity
($\chi^{2}_{\rm g}$), stellar metallicity ($\chi^{2}_{*}$) and both gas-phase and
stellar metallicity $(\chi^{2}_{\rm g}+\chi^{2}_{*})$. The boundary conditions
are adopted to be $0\,\leq\,\lambda\,\leq\,4.5$ and $0\leq\,\omega\,\leq4.5$, respectively.
$\chi^{2}$ contour maps are displayed in Figure\,\ref{fig:f2}. The left, the middle and
the right panels separately plot $\chi^{2}_{\rm g}$, $\,\chi^{2}_{*}$ and
$\chi^{2}_{\rm g}+\chi^{2}_{*}$ contours. The minimum values of $\chi^{2}$ are shown
as red filled asterisks in these three panels of Figure\,\ref{fig:f2}.
The solid lines display the isocontours of
$\Delta\chi^{2}\,=\,\chi^{2}-\chi^{2}_{\rm min}\,=\,$2.3 (blue),
6.17 (cyan) and 11.8 (brown) corresponding to $1\sigma$
(68.3\%), $2\sigma$ (95.4\%) and $3\sigma$ (99.73\%) confidence levels. It should be
emphasized that the oblique line traversing across the red asterisk in the middle panel
arises from the constraint condition $\omega\,\leq\,\lambda+\frac{1}{1-\mu}$ in equation
\ref{eq:metallicityboth} as described in Section \ref{sec:model}.

The left panel of Figure\,\ref{fig:f2} indicates the
degeneracy between parameters $\lambda$ and $\omega$, that is,
the higher inflow rate and the lower outflow rate show similar $\chi^{2}$ values
as the lower inflow rate and the higher outflow rate. The physical
reason is that gas-phase metallicity is diluted by the pristine gas inflow at a fixed
radius, and the enriched outflow process takes a fraction of metals away from
the disc at a fixed radius.
Indeed, the large area of $\chi^{2}$ contours in the left panel indicates that
it is difficult to determine the model parameters only using the gas-phase metallicity.
The corresponding values of the best parameter combinations are
$\lambda\,=\,1.768$ and $\omega\,=\,0.338$.
On the other hand, the middle panel of Figure\,\ref{fig:f2} shows that stellar
metallicity is more sensitive to $\lambda$ than to $\omega$.
The best parameter combinations are $\lambda\,=\,2.456$ and $\omega\,=\,3.485$.
The degeneracy in this case is weaker than the
former case. Moreover, the direction of the degeneracies is different in two cases.
In consequence, the combined fitting is most effective for the constraint of the
parameters.

After using both gas-phase and stellar metallicity
as constraints, the right panel shows that the reasonable range of model parameters
is significantly reduced. The minimum value of $\chi^{2}$ can be found at
$\lambda\,=\,1.994$ and $\omega\,=\,0.192$.
Furthermore, the degeneracy between $\omega$ and $\lambda$ in
the left panel is lifted, which implies that
stellar metallicity may help us to determine the best combination of $\omega$ and $\lambda$.
In other words, the observed stellar metallicity provides an additional constraint on
the chemical enrichment history of M\,101.

We name the model with $\lambda\,=\,1.994^{+0.125}_{-0.136}$ and
$\omega\,=\,0.192^{+0.089}_{-0.088}$ as the best-fitting model of M\,101.
The best-fitting model predicted radial profiles of gas-phase and
stellar metallicity are respectively shown as solid lines
in the left and right panels of Figure\,\ref{fig:f1}. The grey shaded regions in both
panels display the model results within $1\,\sigma$ confidence level, that is,
($\lambda\,=\,1.906, \omega\,=\,0.056$) and ($\lambda\,=\,2.083, \omega\,=\,0.317$).
It should be emphasized that, although we do not take into account the radial
variations of model parameters along the disc in this paper, a good agreement
between the solid lines and the observed data indicates that the
gas outflow rate ($\Psi\,=\,1.418^{+0.089}_{-0.097}\times\rm SFR$) and gas inflow rate
($\Phi\,=\,0.136^{+0.063}_{-0.062}\times\rm SFR$) may reasonably describe the fundamental physical
processes regulating the formation and evolution of M\,101.

Another point we should emphasize is that, the observed metallicity gradient of the gas-phase component
is much steeper than that of stellar component. In other words,  the difference between $Z_{\rm g}$ and
$\langle Z_{\rm \ast}\rangle$ (hereafter $\Delta Z$) decreases with the increase of radius.
Previous studies have shown that there exists a strong correlation between $\Delta Z$ and the mean age of
stellar populations $\langle t\rangle\ $ in the sense that larger $\Delta Z$ corresponds to
older mean stellar age. For further discussion, we refer the reader to \citet{Peng2015} and
to Figure 5 of \citet{Ma2016}. Our results indicate that the inner disc of M\,101 has
an older stellar population than the outer disc, which is consistent
with the inside-out formation scenario of stellar discs.


\begin{figure}
  \centering
  \includegraphics[angle=0,scale=0.65]{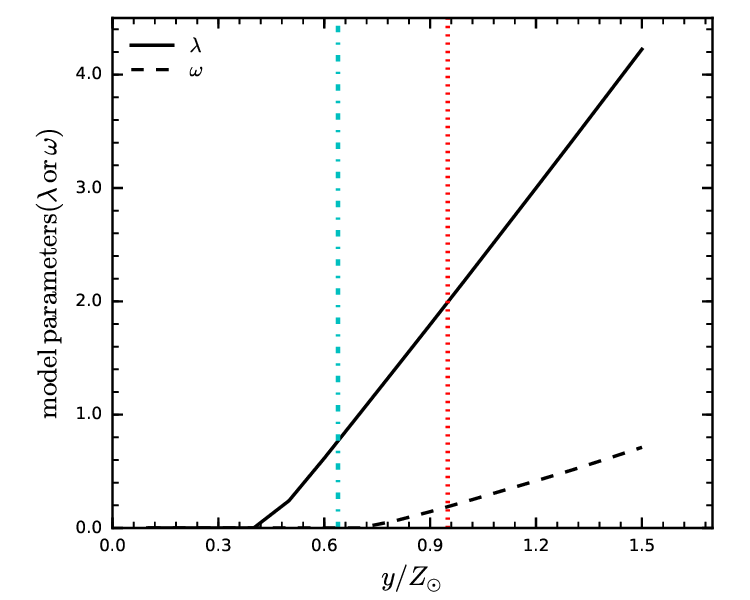}
   \caption{The model parameters $\lambda$ and $\omega$ as a
   function of the yield $y/Z_{\odot}$.
   The solid and dashed lines are corresponding to $\lambda$
   and $\omega$, respectively. The vertical red dotted line denotes
   the value of the yield adopted in this work, while the vertical cyan dash-dotted line
   marks the yield adopted in \citet{Kudritzki2015}.}
  \label{fig:f3}
\end{figure}

Furthermore, we should discuss the influence of the stellar yield  $y$ on the
determination of model parameters. It can be found from equation sets\,\ref{eq:metallicityboth},
\ref{eq:metallicityboth1},
\ref{eq:metallicityoninfall} and \ref{eq:closedbox} in Section \ref{sec:model} that
the stellar yield $y$ is proportional to the resulted metallicity, thus the
adopted $y$ is expected to largely influence the resulting model parameters.
Figure\,\ref{fig:f3} plots the best-fitting model parameters $\lambda$ and $\omega$ as a
function of the yield $y$. The striking feature of Figure\,\ref{fig:f3} is that the value
of outflow parameter $\lambda$ is very sensitive to the adopted yield $y$,
since the main effect of gas outflow process is to take away part of newly
synthesized metals and reduce the metallicity of the ISM.
 In fact, this is a common difficulty in studies of the chemical evolution
of galaxies. Therefore, the absolute value of $\lambda$ derived in this paper is not robust, and
we can only estimate the relative probability of $\lambda$
for given stellar yield.

Finally, we note that the best combination of $\lambda$ and $\omega$
for M\,101 derived by \citet{Kudritzki2015} is $\omega\,=\,0.0$ and $\lambda\,=\,0.98$,
that is, $\Phi\,=\,0$ and $\Psi\,=\,0.98$ in units of SFR, who use the gas-phase
metallicity to constrain the model. The smaller values of $\lambda$ and $\omega$ for
M\,101 in \citet{Kudritzki2015} than ours (see both left and right
panels of Figure\,\ref{fig:f2}) mainly due to the fact that
they adopted a smaller stellar yield than ours.

\section{Summary}

In this work, the radial distribution of gas-phase and stellar metallicity
of star-forming galaxies is investigated by means of an analytical chemical evolution
model. We find that the gas-phase and stellar metallicity can be derived
by the ratio of stellar-to-gas mass surface densities.
Through comparing the gas-phase metallicity with the stellar metallicity as a
function of the stellar-to-gas mass ratios with models
of different combinations gas inflow and and outflow rates, it
is shown that both gas inflow and outflows can
reduce the gas-phase metallicity, but there exists degeneracy effect between
$\omega$ and $\lambda$. On the other hand, stellar metallicity is more
sensitive to $\lambda$ than to $\omega$, and this helps
to reduce the
degeneracy effect. The analytical chemical evolution model
is applied to the nearby disc galaxy M\,101. By means of the classical
$\chi^{2}$ methodology, $\omega$ and $\lambda$ are better determined
by simultaneously using gas-phase and stellar metallicity as the observed
constraints, which further indicates that stellar metallicity is an
important additional observable to constrain the SFH of star-forming galaxies. Our results
also show that relatively strong gas outflows but weak inflows occurred
on the disc of M\,101 during its evolutionary history.

Recent IFU surveys provide large samples of data
for star-forming galaxies which include spatially resolved information of observed
stellar mass, gas mass, gas-phase metallicity and stellar metallicity. This will
provide an opportunity for further tests our method.
We plan to apply the analytical chemical evolution model to a large sample of star-forming
galaxies to constrain the gas inflow and outflows during their evolutionary histories
in our future work.

\section*{Acknowledgements}

We thank the anonymous referee for thoughtful comments and insightful
suggestions that greatly improved the quality of this paper.
This work is supported by National Key R\&D Program of China (No. 2019YFA0405501).
Xiaoyu Kang and Fenghui Zhang are supported by the National
Natural Science Foundation (NSF) of China (No. 11973081, 11573062,
11403092, 11390374, 11521303), the YIPACAS Foundation (No. 2012048),
the Chinese Academy of Sciences (CAS, KJZD-EW-M06-01),
the NSF of Yunnan Province (No. 2019FB006)
and the Youth Project of Western Light of CAS.
Ruixiang Chang is supported by the National NSF of China
(No. 11373053, 11390373). Rolf Kudritzki acknowledges
support by the Munich Excellence Cluster Origins
Funded by the Deutsche Forschungsgemeinschaft
(DFG, German Research Foundation) under the German
Excellence Strategy EXC-2094 390783311.


\section*{Data availability}

The data underlying this article will be shared on
reasonable request to the corresponding author.











\bsp	
\label{lastpage}
\end{document}